# ChatGPT for Generating Questions and Assessments Based on Accreditations

Rania A Aboalela

[1]Department of Information Systems, King Abdulaziz University, Jeddah, Kingdom Of Saudi Arabia
raboalela@kau.edu.sa

## Abstract

*This research aims to take advantage of artificial intelligence techniques in producing students' assessment that is compatible with the different academic accreditations of the same program. The possibility of using ChatGPT technology was studied to produce an academic accreditation-compliant test NCAAA and ABET. A novel method was introduced to map the verbs used to create the questions introduced in the tests. The method allows a possibility of using the ChatGPT technology to produce and check the validity of questions that measure educational outcomes. A questionnaire was distributed to ensure that the use of ChatGPT to create exam questions is acceptable by the faculty members, as well as to ask about the acceptance of assistance in validating questions submitted by faculty members and amending them in accordance with academic accreditations. The questionnaire was distributed to faculty members of different majors in the Kingdom of Saudi Arabias' universities. 120 responses obtained with 85% approval percentage for generate complete exam questions by ChatGPT. Whereas 98% was the approval percentage for editing and improving already existed questions.*

## 1. Introduction

In this research, we found a method for faculty members to benefit from using ChatGPT technology. Accordingly, the teacher can ask the application to give questions according to the teacher's request and on condition that they are compatible with the required academic accreditation. As well as producing questions that correspond to more than one provision and different accreditations. In this research, Bloom taxonomy was used to find a map between the requirements of the NCAAA [1] & ABET [2] to produce questions that meet the requirements of the NCAAA & ABET. Using Bloom Taxonomy for mapping between the verbs of the same curriculum is a matter that has been studied a lot and contributed to increasing the quality of the questions presented to measure students' knowledge [3] [4] [5]. In this research Bloom Taxonomy was used to map between the various academic accreditations. Then an innovative method was developed to produce questions related to common requirements using ChatGPT. Mapping was made between the requirements and conditions of the different academic accreditations for the questions that measure the knowledge of the students. The mapping was found through the Bloom Taxonomy verbs which measure different Levels of Knowledge.

With mapping between the requirements of different academic accreditations in creating questions that measure students' knowledge by requiring certain verbs, we can use the required verbs questions through accreditations. It is the key to using technology to create questions compatible with academic accreditations accurately and quickly, which saves instructor time, as well as making it easier for reviewers to confirm the eligibility of programs approved .The faculty members' willingness to use this technology to speed up and facilitate the production of questions was studied. As well as the extent of satisfaction with the map between the verbs of the outcomes of ABET with zones of NCAAA three-level domains. A questionnaire was distributed to faculty members to measure the extent to which members accept the use of ChatGPT technology to produce appropriate questions according to the curriculum being taught and according to the educational outcomes of the course. The response was from 120 faculty members. 85% of those wishing to use this technique in creating questions were obtained. 98% of those wishing to benefit from this technology only help, guide and correct in creating questions.

## 2. LITERATURE REVIEW

The Over the last few years, the area of Natural Language Processing (NLP) has grown significantly. However, the introduction of ChatGPT (Chat Generative Pre-Trained Transformer) has rekindled interest and excitement in this technology [6], ChatGPT was developed by Open AI and distributed to the public in November 2022. ChatGPT grew so quickly that it reached one million users in five days, whereas Facebook took 300 days, Twitter took 720 days, and Instagram took 75 days [7].

ChatGPT is a big language model with amazing comprehension and production skills that closely match human speech. Its excellent ability at answering queries, engaging in conversations, and providing logical and contextually relevant replies has marked an important milestone in the progress of conversational AI. The multifarious applications of ChatGPT and its potential to enhance user productivity across diverse industries have ignited fresh deliberations surrounding this state-of-the-art artificial intelligence (AI) technology [8].

Regardless of its advanced capabilities, ChatGPT is essentially a complex chatbot with origins in the early stages of Long Short-Term Memory (LSTM) development [9] and has been widely used in a variety of fields such as automated customer service support, E-Commerce, Healthcare, and Education. An area of artificial intelligence (AI), machine learning refers to the capacity of computer systems to learn from experience without being explicitly programmed. Deep learning has developed as a high-performing prediction tool with developments in processing power, increased data availability, and algorithmic improvements [10], [11], [12], [13]. In addition, for ChatGPT for assessment [14]. ChatGPT is an enormous language model with amazing comprehension and production skills that closely match human speech. Its outstanding performance in answering questions, engaging in conversations, and generating coherent and contextually appropriate replies has marked a key milestone in the evolution of conversational AI [9]. The first GPT model, GPT-1, was introduced in 2018, followed by GPT-2 in 2019 and GPT-3 in 2020. Each successive iteration of the model has improved model size, training data, and performance on language tasks. On 30 November 2022, Open AI released a free preview of ChatGPT, their latest AI chatbot, increasing OpenAI's projected worth to US$29 billion [15]. A chatbot is a software program powered by artificial intelligence that can have human-like conversations. Users can ask inquiries or make requests, and the system will answer within seconds. ChatGPT gained one million users only five days after its original debut [16].

## 3. CHATGPT IN HIGHER EDUCATION ASSESSMENT

Teaching and learning are complex processes, taking place in various settings and different forms. These processes, which are rarely assessment-free, have been largely affected by technological inventions which include changes in AI-based teaching strategies and instructional materials [17].The recent development of ChatGPT has shook every academic institution, and while we are still learning about its full potential and pitfalls, it is worthwhile to provide an introductory perspective. Due to the exceptional powers of the chatbot's human-like capabilities that transcend most current technologies that we have encountered, there are unprecedented opportunities for ChatGPT in academics [18]. ChatGPT has received remarkable interest from the academic community and the public in recent months. Because it is unlikely that the chatbot was designed with the intention of serving as a substitute for academic writing, its application to academic writing is a byproduct of artificial intelligence (AI) brilliance. AI applications in language learning and teaching have increased in parallel to new learning models and modes. Recently, ChatGPT has directed attention towards more reliable and valid assessment tools that gauge learning outcomes [17]. If given the choice, students across the world would find a method to avoid tests, therefore we are all concerned that, despite its benefits, some students may misuse it. While academia is far from experiencing an assessment integrity crisis, the rise of sophisticated AI and technologies that might facilitate cheating cannot be overlooked. some of us feel that certain epistemic implications exist for the efficacy of ChatGPT in evaluations; yet possible dangers would not signal the end of our resolve. So far, we know that some university programs are more vulnerable for example, Management Studies and Information Technology, but educators are not new to academic cheating - they simply don't fully comprehend ChatGPT yet. Despite its inescapable usage in some academic contexts, no compelling rationale was found to support its use in evaluations. Students are not taught to "copy and paste," but rather to "think and write critically." It should thus be of worry that ChatGPT has passed medical school examinations [19] and MBA evaluations. Utilitarian ethicists will see no reason to oppose the AI revolution, even if it reduces the veracity of higher education assessment; however, consequentialists will argue that the spread of AI and questions about the ethics of their use will shape the future of research in many areas, including the long-term purpose and utility of higher education [20]. With the growing volume of medical data and the complexity of clinical decision-making, NLP technologies might theoretically aid clinicians in making timely, informed judgements, improving overall healthcare quality and efficiency. Without any specialized training, ChatGPT performed at or near the passing criteria for the United States Medical Licensing Exam (USMLE), indicating its potential for medical education and clinical decision assistance. Furthermore, technological improvements have resulted in the democratization of knowledge, with patients no longer relying primarily on healthcare providers for medical information. Instead, patients are increasingly turning to search engines and, more recently, artificial intelligence chatbots as handy and accessible sources of medical information. ChatGPT and other recently released chat-bots engage in conversational interactions and provide authoritative-sounding responses to complex medical queries. However, despite its potential, ChatGPT frequently produces seemingly credible but incorrect outputs, necessitating caution in its applications in medical practice and research. These engines' dependability and correctness have not been evaluated, particularly in the context of open-ended medical queries that clinicians and patients are likely to ask [21]. The study conducted by Aidan Gilson on How Does ChatGPT perform on the United States Medical Licensing Examination in 2023 stated that ChatGPT performing at a greater than 60% threshold on the NBME-Free-Step-1 data set, shows that the model achieves the equivalent of a passing score for a third-year medical student. Additionally, ChatGPT's capacity to provide logic and informational context across most answers has been proved. These facts taken together make a compelling case for the potential applications of ChatGPT as an interactive medical education tool to support learning [22].

## 4. ABET & NCAAA

ABET is a form of quality assurance for programs in the areas of applied and natural science, computing, engineering, and engineering technology [2]. ABET accreditation is recognized globally for providing assurance that a college or university program meets the quality standards of the profession for which that program prepares graduates. ABET requirements are applied in the undergraduate program of the Department of Information Systems in the College of Computing and Information Technology at King Abdulaziz University, Rabigh Branch. The student outcome (SO) of the Website Design and Development course (COIS 492) was used, then those outcomes were simulated to be applied in producing learning outcomes that comply with the requirements of the NCAAA [1], which is the national academic accreditation for educational programs in the Kingdom of Saudi Arabia. The learning outcomes for students are measured based on action verbs that are used to create questions that It measures the students' skills and knowledge of the subject. Then, the students' knowledge is measured based on the grades that the student obtains for each question related to a specific knowledge or skill. Obtaining national and international accreditation for one academic program is applicable in Saudi universities. It will enable one academic program to obtain different academic accreditations through several methods. There are many studies demonstrating a possible link between ABET and the NCAAA in the Kingdom of Saudi Arabia. Some research that has been published to prove these possibilities [23] [24]

## 5. MERGE NCAAA WITH ABET

The question verbs that are used to measure ABET's educational outcomes are key to making the connection with the NCAAA requirements, which also requires question verbs to measure educational knowledge and skills. In addition to subdividing the areas of education in the NCAAA into three domains: Knowledge, Skills and Values. Each domain has its own verbs that measure it. Hence, a common link between ABET and NCAAA specific educational outcome requirements has been found. They are the same verbs that measure educational outcomes and were used in the production of automatic questions through ChatGPT.

## 6. THE ROLE OF BLOOM TAXONOMY.

The role of Bloom Taxonomy in assessment of knowledge [25] [26] [27] is seen in mapping the special verbs to measure educational outcomes related to ABET accreditation as well as to NCAAA accreditation, which can make those standard verbs that measure a specific educational outcome the same used in questions that meet the requirements of ABET as well as the requirements of the NCAAA. Thus, when assuming a common question verb to measure the same educational outcome that is measured by this question. The question now left to do is to divide the verbs used into the areas whose measurement requires NCAAA accreditation. Hence the role of Bloom Taxonomy through which verbs that measure skills and knowledge are divided into six domains. They are as Figure 1. The six Bloom Taxonomy domains have been subdivided into NCAAA-specific fields. The higher five most complex domains Applying, analyzing, evaluating, and creating, have been reduced to the NCAAA domain of skills. The lower most complex domain Understanding remains the subject area of knowledge of the NCAAA domain. As for the verbs, they were divided according to the verbs of Bloom's taxonomy. Figure Shows the mapping. Once you get the map between taxonomy verbs with the NCAAA, you can find measure verbs that measure a specific skill, either according to the required verbs from the point of view of accreditation of ABET or accreditation of the NCAAA.

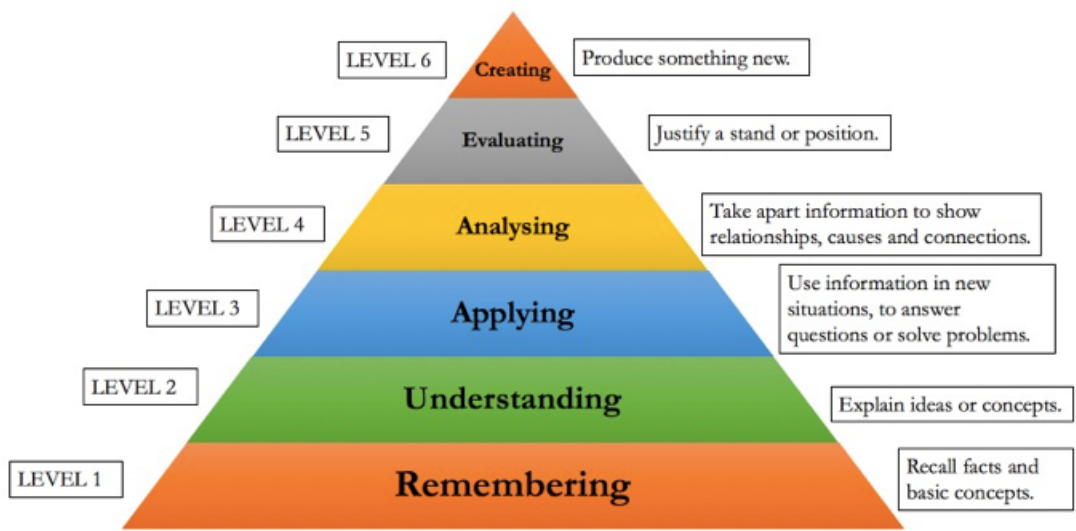

Figure 1: The Bloom Taxonomy Verbs [28]

## 6.1. ABET 6 Students' Outcomes of Information System Bachelor Program.

The Bloom Taxonomy verbs would be mapped to the ABET six SO's for an Information Systems program at King Abdulaziz University in the College of Computing and Information Technology in Rabigh as the following:

1. Analyze a complex computing problem and to apply principles of computing and other relevant disciplines to identify solutions. SO 1 corresponds to the third level verb in Bloom's Taxonomy

2. Design, implement, and evaluate a computing-based solution to meet a given set of computing requirements in the context of the program's discipline. SO 2 corresponds to the level 6, 5 and 4 based on the SO subpoint, and the action verb in Bloom's Taxonomy

3. Communicate effectively in a variety of professional contexts. SO 3 corresponds to special verb in Bloom's Taxonomy based on the asked question target and to the value domain of NCAAA

4. Recognize professional responsibilities and make informed judgments in computing practice based on legal and ethical principles. SO 4 corresponds to special verb in Bloom's Taxonomy based on the target of the asked question and to the value domain verb in NCAAA

5. Function effectively as a member or leader of a team engaged in activities appropriate to the program's discipline SO 5 corresponds to special verb in Bloom's Taxonomy based on the asked question target and to the value domain verb in NCAAA

 (Affective Learning)

6. Support the delivery, use, and management of information systems within an information systems environment. SO 6 corresponds to lower level (Understanding) of Bloom's Taxonomy based on the asked question target and to the SKILL domain verb in NCAAA. For each SO there are special verbs for the one director who must ask the question. To find a method to connect the verbs of each ABET'S SO with the NCAAA'S domains, the six ABET outcomes were divided into the six Bloom Taxonomy domain. , which then will be divided into the three NCAAA. Figure 2 shows the mapping between Bloom taxonomy & NCAAA. The table 1 summarizes the mapping between the accreditations ABET & NCAAA and the Bloom Taxonomy.

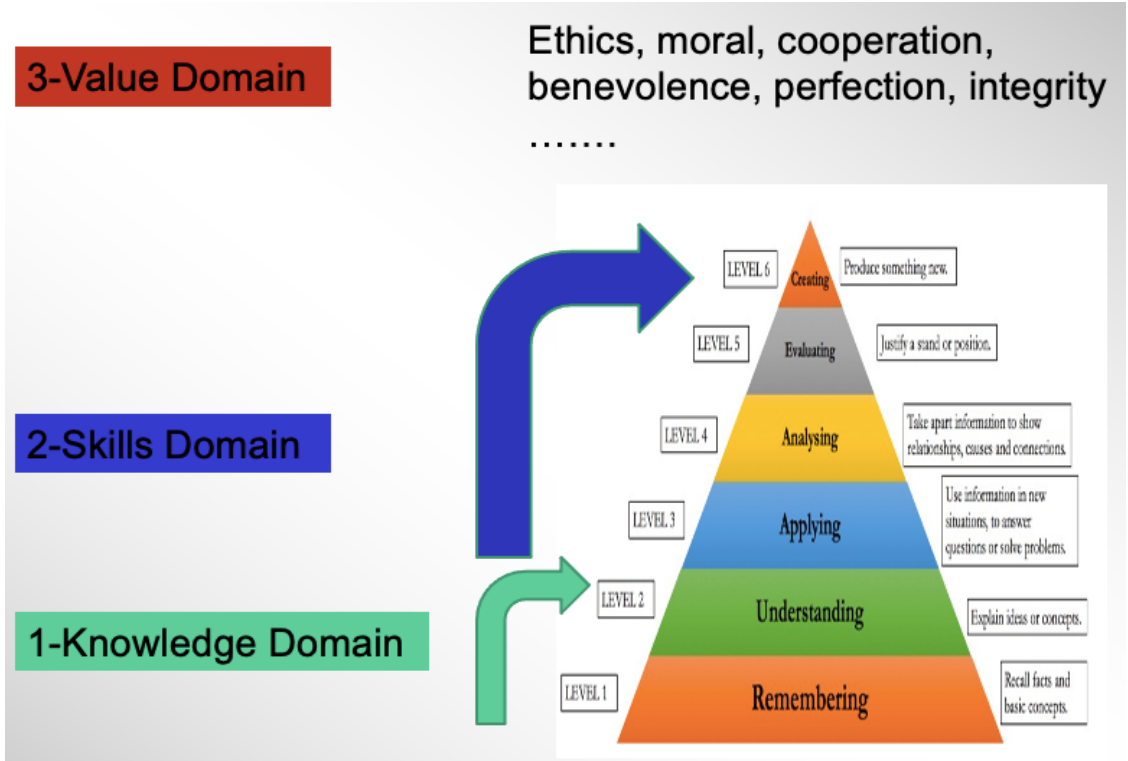

Figure 2: Mapping Bloom Taxonomy to NCAAA three domains

Table 1: The Mapping Between the Accreditations and the Bloom Taxonomy.

| ABET SO | NCAAA Domains | Bloom Taxonomy Level |
|---------|---------------|----------------------|
| SO1 | Skills domain | L3, L5, L6 |
| SO2 | Skills domain | L5, L6 |
| SO3 | Value domain | L2, L3 |
| SO4 | Value domain | L2, L3 |
| SO5 | Value domain | L2, L3 |
| SO6 | Skills Domain | L2, L3 |

**6.2. Mapping Question Verb to ABET & Bloom Taxonomy.**

ABET specializes special verbs ask for each SO. This feature makes it easier to create questions that are compatible with accreditations. Whereas the NCAAA requires the use of certain verbs, these verbs are the key to automatic question generation and validation using ChatGPT technology. Where the machine is fed and trained using generative AI, it is easy to use ChatGPT technology. These verbs can be used to create questions related to NCAAA. They will be jointly meeting ABET & NCAAA requirements. Thus, the questions created are compatible with both accreditations if the academic program obtains two different accreditations Table 2 shows the question verbs which specified by ABET to measure each SO. The table also shows the bloom taxonomy mapped verb. These questions verbs are supposed to be the same for ABET & NCAAA.

Table 2. Question Verbs Mapped to NCAAA & ABET SO

| Question Verb | Bloom Taxonomy VERB LEVEL | The ABET SO # |
|---|---|---|
| Appraise, assess, evaluate, compare, contrast, criticize, differentiate, discriminate, distinguish, examine, experiment, question, test | [Analyzing] | 1.1 |
| Choose, demonstrate, employ, illustrate, interpret, operate, schedule, sketch, draw, solve, use, write. | [Applying] | 1.2 |
| Assemble, construct, create, design, develop, formulate, write . | [Creating] | 2.1 |
| Choose, demonstrate, employ, illustrate, interpret, operate, schedule, sketch, draw, solve, use, write | [Applying] | 2.2 |
| [ Affective Learning] Appreciate, accept, attempt, challenge, defend, dispute, join, judge, justify, question, share, support . | [Evaluating] | 2.3 |
| Choose, demonstrate, employ, illustrate, interpret, operate, schedule, sketch, draw, solve, use, write. | [Applying] | 3.1-3.3 |
| Classify, describe, discuss, explain, identify, locate, recognize, report, select, translate, paraphrase | [Understanding] | 4.1-4.3 |
| Choose, demonstrate, employ, illustrate, interpret, operate, schedule, sketch, draw, solve, use, write. | [Applying] | 5.1 |
| [ Affective Learning] Appreciate, accept, attempt, challenge, defend, dispute, join, judge, justify, question, share, support | Any verb level which should be determined by the SO of the topic. | 5.2-5.3 |
| Appraise, assess, evaluate, compare, contrast, criticize, differentiate, discriminate, distinguish, examine, experiment, question, test | [Analyzing] | 6.1 |
| Choose, demonstrate, employ, illustrate, interpret, operate, schedule, sketch, draw, solve, use, write. | [Applying] | 6.2 |
| Classify, describe, discuss, explain, identify, locate, recognize, report, select, translate, paraphrase. | [Understanding] | 6.3 |

Whereas the subpoints 1.1, 1.2, 2.1, 2.2, 2.3, 3.1-3.3, 4.1-4.3, 5.1, 5.2-5.3, 6.1, 6.2 and 6.3 are subpoints of the major 6 SO and they measure the following SO.

1.1: An ability to Analyze a complex computing problem ( Analyzing)

1.2: An ability to Apply principles of computing and other relevant disciplines to identify solutions (Applying)

2.1: An ability to design a computer-based system, process, component, or program to meet desired needs.

2.2: An ability to implement a computer-based system, process, component, or program to meet desired needs.

2.3: An ability to evaluate a computer-based system, process, component, or program to meet desired needs.

3.1: An ability to conduct an oral presentation using effective communication skills. (Applying)

3.2: An ability to write in a clear, concise, grammatically correct and organized manner. (Applying)

3.3: An ability to develop appropriate illustrations including hand sketches, computer generated drawings/graphs and pictures. (Applying)

4.1: Understanding of professional responsibilities, ethical theories, legal and social issues. (Understanding)

4.2: Understanding of cyber security threats and corresponding procedures to mitigate these threats. (Understanding)

4.3: Understanding of risk management, security policies and audit procedures. (Understanding)

5.1: An ability to prepare a work schedule for the assigned task and complete it within the appropriate deadlines. (Applying)

5.2: An ability to participate in team meetings with full preparedness for providing useful input. (Affective Learning)

5.3: An ability to share ideas among the team and promote good communication among the team members. (Affective Learning)

6.1: Support the delivery of information systems within an information Systems environments.

6.2: Support the use of information system within an information Systems environments.

6.3: Support the management of Information Systems within an information Systems environments.

Studying the extent to which faculty members accept the use of ChatGPT technology. In order to produce student assessment questions and tests. These subpoints are extracted from ABET official documents from IS department of FCIT of King Abdulaziz University [29].

### 6.3. ChatGPT for Generating Questions

A special application will be created for each curriculum using ChatGPT. The databases are fed with the content of each curriculum. Training the machine to produce questions and then predict the questions at the most appropriate level according to the requirements and conditions of each accreditation, then linking them together as studied in this research. The question here is to what extent iis possible to deduce questions so that it is possible to ask questions outside the curriculum to stimulate students to think. In this case we suggest taking the advantage of deep learning & Generative AI applications. Specified verb measure specified Skills or Knowledge will be determined by the coordinator or the instructor of the course at any field. How does it work?

The instructor asks the machine to generate questions related to a specific topic of the curriculum. The application creates questions that are compatible with the accreditations required by the educational institution, such as the NCAAA accreditation required in educational institutions in Saudi academies in the Kingdom of Saudi Arabia, in addition to ABET accreditation. This is due to the existence of mapping between the requirements of the two accreditations, which allows the work of unified and approved questions that are compatible with the two different accreditations (National & International). The application is asked to produce a set of questions from a specific topic and determines the number of questions and conditions so that they are identical to the required accreditations. The application is also possible to benefit from the application by modifying the questions that were produced by the same teacher or any official to achieve the required quality conditions.

### 6.3.1. Example:

From COIS492 Web Design & Developement course which From the Department of Information Systems at the University, College of Computing and Information Technology, Rabigh Branch. The course satisfies three SO's of ABET. Student Outcome 6, Student Outcomes 4 and 2. The instructor asks the app about the subject Web Design and development, could be specified to create limited questions for specific curriculum area, an HTML topic that achieves Student Outcome 6.2, Student Outcome 4.1 and 2.2. For example, question that asks about the html topic specifies asking to write a code. The question would be **write** a code shows the output of seven lines on the screen. The verb (write) is in the list of measure verbs relevant to ABET SO 2.1 which is 2.1 An ability to design a computer-based system, process, component, or program to meet desired needs. This verb related to Bloom Taxonomy level six domain which is **(Creating)** which is considered as skills based on NCAAA. In other word the verb write is compatible with the verbs of ABET's Learning Outcome 6.2 which correspond to Bloom Taxonomy Level 6 **Creating** which in turn is in a **skills** domain according to the NCAAAA accreditation domains. Accordingly, with questions that comply with the terms of accreditations. As a question of the Students Outcomes SO 2.1. Similarly for the rest of the students' students' outcomes. instead of creating the question by the instructor the application uses ChatGPT creates the questions will be innovative, varied, and in line with the course objective. In addition, it would be saving the time. A questionnaire was conducted for faculty members regarding to measure the desirability of creating questions through ChatGPT technology.

## 7. QUESTIONNAIRE:

The evaluator's satisfaction at the Saudi Universities was measured about their acceptance of the use of ChatGPT technology to produce questions for the subjects they teach. It was also asked if they accept correcting and modifying the questions created by the teachers to comply with educational quality standards and requirements for academic accreditation. Two questions were presented as the following. The first question was: Do you support the use of ChatGPT technology to produce assessment questions and tests?
The second question was as: Do you support the use of ChatGPT technology. In assisting in conducting tests (error correction, guidance), in compliance with the requirements and conditions of accreditations. The questionnaire was distributed to a group of Saudi universities in various colleges and majors.

## 7.1. The Result

The questionnaire was distributed randomly to faculty members through whatsup technology. The response was from 120 faculty members. 85% of those are wishing to use this technique in creating questions were obtained. 98% of those wishing to benefit from this technology only help, guide and correct in creating questions. Figure 3 & 4 shows the result. The percentage of those who rejected the use of this technique in creating questions was 15 percent. On the other hand, the refusal was in the creation and the support in helping only in the work and the amendment is 98%..

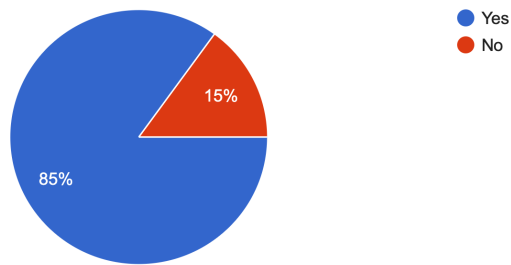

Figure 3. The percentage of acceptance of ChatGPT generates questions.

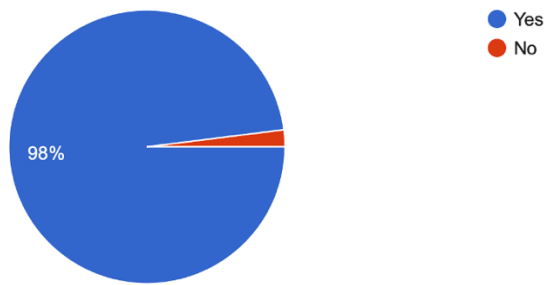

Figure 4. The percentage of acceptance of ChatGPT guides the evaluator to generate questions.

# 8. THE CONCLUSION & FUTURE WORKS

In this research, a novel method was presented to produce questions that measure the learning outcomes of a single course using the ChatGPT technology. In addition to the compatibility of these questions with the requirements of different academic accreditations. In this research, a method was found to link the verbs that are conditioned in the NCAAA accreditation and the ABET accreditation to measure the educational outcome of the student. Through this mapping, we were able to find a method to produce questions that are compatible with accreditations and achieve high quality and efficiency. In the upcoming works, the application will be created, the extent of its purpose. The use of this technology will be expanded in the production of questions and the quality of education through the ChatGPT technology